\documentclass[aps,prb,showpacs,twocolumn]{revtex4-2}
\usepackage{amssymb}
\usepackage{amsmath}
\usepackage{graphicx}
\usepackage{epsfig}
\usepackage{subfigure}
\usepackage{color}

\begin{document}

\title{Hilbert Space Fragmentation in Hardcore Bose and Fermi Hubbard Models on
Generalized Lieb Lattices}
\author{D. K. He}
\author{Z. Song}
\email{songtc@nankai.edu.cn}
\affiliation{School of Physics, Nankai University, Tianjin 300071, China} 

\begin{abstract}
We study the Hilbert space fragmentation (HSF) in hardcore Bose and Fermi
Hubbard models in the framework of the restricted spectrum generating
algebra (RSGA). We present a family of hardcore Bose-Hubbard models with
repulsive density-density interactions on a generalized Lieb lattice. We
show that this system possesses the RSGA structure in the large interaction
strength limit, exhibiting quantum HSF. It allows us to construct a set of
exact condensate eigenstates, possessing off diagonal long-range order.
Based on numerical simulations conducted on several representative lattices,
we demonstrate the existence of weak fragmentations when the constraints are
not exact. As applications, we also studied the connection between HSF and
RSGA in modified fermionic Hubbard models, where the $\eta $-pairing
states are shown to be energy towers, acting as quantum scars.
\end{abstract}
\maketitle
\section{Introduction}

The eigenstate thermalization hypothesis (ETH) not only explains
thermalization in isolated systems within the framework of quantum mechanics 
\cite%
{Deutsch_Quantum,srednicki1994chaos,d2016quantum,borgonovi2016quantum,gogolin2016equilibration,serbyn2021quantum}%
, but also seems to pose challenges for quantum simulation and quantum
information tasks. Fortunately, evidence shows that the ETH can be violated
in some situations \cite{nandkishore2015many,abanin2019colloquium,bernien2017probing,choi2019emergent,turner2018weak,pai2019localization,kwan2025minimal,feng2022hilbert,zhao2020quantum,turner2021correspondence,mukherjee2020collapse,van2020quantum,bluvstein2021controlling,surace2021exact,yang2025constructing}
. Most eigenstates still follow the ETH, yet non-thermal behavior can be
observed when the system is prepared in some special initial states. A
promising mechanism for the anomalous thermalization is the Hilbert space
fragmentation (HSF). It originates from intrinsic kinetic constraints \cite%
{schecter2019weak,yang2020hilbert,moudgalya2022hilbert,li2023hilbert,francica2023hilbert,nicolau2023flat}%
, which fragment the Hilbert space into dynamically isolated subspaces,
thereby rendering some states inaccessible and preventing full
thermalization. Constrained models, such as the PXP model \cite%
{lesanovsky2012interacting,turner2018weak}, constrained spin chains \cite%
{lingenfelter2024exact}, and dipole-conserving hopping models \cite%
{sala2020ergodicity}, were the first to exhibit fragmented dynamics, which
is indicative of HSF. In systems with fragmented Hilbert spaces, certain
subspaces may contain special eigenstates that are the quantum many-body
scars (QMBS) \cite%
{shiraishi2017,moudgalya2018,moudgalya20182,khemani2019,ho2019,shibata2020,mcclarty2020,richter2022,jeyaretnam2021,turner2018weak,turner20182,shiraishi2019,lin2019,choi2019emergent,khemani2020,dooley2020,dooley2021}
. These non-thermal states are typically embedded within the bulk spectrum
of the system and span a subspace in which initial states fail to thermalize
and instead exhibit periodic behavior. In practice, the kinetic constraint
is usually not imposed naturally, but induced from the particle-particle
interactions. Consequently, the corresponding interaction strength
determines the degree of the HSF, which then influences the formation of the
scar.

Besides the development of the theory, concrete examples are beneficial for
understanding the mechanism of HSF. A growing body of models has recently
been shown to host QMBS, prompting attempts to subsume them within unified,
systematic frameworks \cite{shiraishi2017,mark2020unified,moudgalya2020eta,pakrouski2020many,ren2021quasisymmetry,o2020tunnels}. Among them, the restricted spectrum
generating algebra (RSGA) formalism introduced in Ref. \cite{moudgalya2020eta} provides a
classification of QMBS that lies at the focus of this work. It reveals the
features and structure of a class of Hamiltonians that possess an exact energy tower. Recently, the $\eta $-pairing state \cite{yang1989eta,yang1990so} has received
a renewed interest from a perspective of HSF \cite{moudgalya2020eta,vafek2017entanglement,Mark20,pakrouski2020many,Pakrouski21}. The modified
Hubbard models are proposed to meet the condition that the $\eta $%
-pairing state remains an eigenstate but is not protected by the $\eta $%
-pairing symmetry.

In this paper, we investigate the connection between HSF and QMBS through a
family of hardcore Bose-Hubbard models with repulsive density-density
interactions on a generalized Lieb lattice. We show that this system
possesses the RSGA \cite{moudgalya2020eta} structure under a kinetic
constraint. This constraint can be achieved in the large interaction
strength limit, leading to HSF. The RSGA allows us to construct a set of
exact condensate eigenstates that possess off-diagonal long-range order.
Numerical simulations are conducted on several representative lattices with
different interaction strengths. The results demonstrate the existence of
weak fragmentation when the constraints are not exact. In addition, we
investigate such fragmentation in a fermionic system. As applications, we
also studied the connection between HSF and RSGA in modified fermionic
Hubbard models, where the $\eta $-pairing states are shown to be energy
towers, acting as quantum scars. The advantage of this model is that the
kinematic constraints can be realized naturally due to the statistics of
fermions. Our work provides an explicit relationship between a model
featuring interaction-induced constraints and the construction of energy
towers.

The structure of this paper is as follows. In Sec. \ref{Model with kinematic constraints and RSGA}, we introduce the model Hamiltonian and show that it
possesses RSGA under the kinematic constraints. In Sec. \ref{Weak Hilbert space fragmentation}, we investigate the impact of relaxing these
constraints on the energy tower structure. In Sec. \ref{Connection to Fermi Hubbard model}, we apply the result to the Fermi systems. Finally, in Sec. 
\ref{Summary}, we provide a summary and discussion.

\section{Model with kinematic constraints and RSGA}

\label{Model with kinematic constraints and RSGA}

\begin{figure}[t]
\centering
\includegraphics[width=0.5\textwidth]{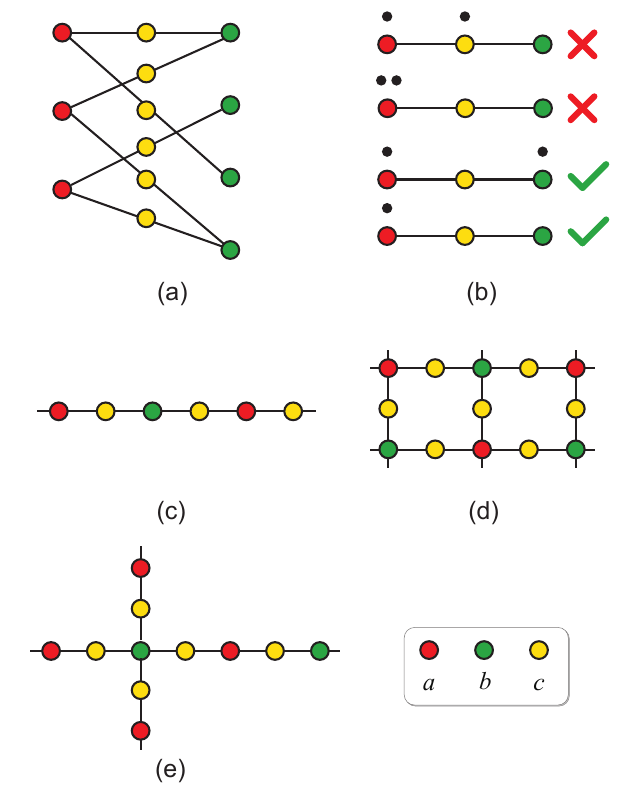}
\caption{Schematic illustrations for the features of the studied system. (a)
A generalized Lieb lattice, consisting of three sub-lattices A, B, and C,
that are denoted by red, yellow, and green-filled circles, respectively. The
black lines denote the connections between sublattice A and B. Each
connection corresponds to a single site of the sub-lattice C. The two
nearest-neighbor hopping strengths along a connection are real and identical.
(b) The conditions for the RSGA on the configurations of the boson filling
specify that the doubly occupied and nearest-neighbor pair states are
forbidden. (c) and (d) are two examples which represent 1D and 2D Lieb
lattices.}
\label{fig1}
\end{figure}

Considering a hardcore Hubbard model with infinite nearest-neighbor (NN) on
a generalized Lieb lattice, the Hamiltonian has the form

\begin{eqnarray}
H &=&\sum_{i,j=1}^{N_{a},N_{b}}\kappa _{ij}[\left( a_{i}^{\dagger
}+b_{j}^{\dagger }\right) c_{i,j}+\text{\textrm{H.c.}}]  \notag \\
&&+V\sum_{i,j=1}^{N_{a},N_{b}}\left( a_{i}^{\dagger }a_{i}+b_{j}^{\dagger
}b_{j}\right) c_{i,j}^{\dagger }c_{i,j}\text{\textrm{,}}  \label{H}
\end{eqnarray}%
where $\left\{ a_{l}\right\} $,$\ \left\{ b_{l}\right\} $, and $\left\{
c_{ij}\right\} $ are hardcore boson annihilation operators on three sets of
lattices $A$, $B$ and $C$, with the lattice numbers $N_{a}$, $N_{b}$, and $%
N_{c}$, respectively. Here, $\left\{ \kappa _{ij}\right\} $ is a set of
real numbers, representing the hopping strengths. The nonzero $\kappa _{ij}$
determines the existence of the lattice site at $\left( i,j\right) $. The
total number of nonzero $\kappa _{ij}$ is then $2N_{c}$. We note
that when the two lattices $A$ and $B$ constitute a square lattice, the
whole lattice is a standard 2D Lieb lattice. The total number of particles is
conserved, i.e.,%
\begin{equation}
\lbrack \sum_{i,j=1}^{N_{a},N_{b}}\left( a_{i}^{\dagger
}a_{i}+b_{j}^{\dagger }b_{j}+c_{i,j}^{\dagger }c_{i,j}\right) ,H]=0.
\end{equation}%
A schematic illustration of the generalized Lieb lattice is presented in
Fig. \ref{fig1}.

We introduce a set of operators%
\begin{eqnarray}
\eta ^{+} &=&\left( \eta ^{-}\right) ^{\dag
}=\sum_{i=1}^{N_{a}}a_{i}^{\dagger }-\sum_{j=1}^{N_{b}}b_{j}^{\dagger }, \\
\eta ^{z} &=&\frac{1}{2}\left[ \sum_{i=1}^{N_{a}}\left( 2a_{i}^{\dagger
}a_{i}-1\right) -\sum_{j=1}^{N_{b}}\left( 2b_{j}^{\dagger }b_{j}-1\right) %
\right] ,
\end{eqnarray}%
they are pseudo-spin operators, satisfying the su(2) algebra

\begin{eqnarray}
\left[ \eta ^{+},\eta ^{-}\right] &=&2\eta ^{z}, \\
\left[ \eta ^{z},\eta ^{\pm }\right] &=&\pm \eta ^{\pm }.
\end{eqnarray}%
We note that the operator $\eta ^{+}$ is the essential $\eta $-pairing
operator obtained by replacing the on-site pair-fermionic operator with a
hard-core bosonic operator. Further discussion will be given in Sec. \ref%
{Connection to Fermi Hubbard model}. A straightforward derivation shows the
following conclusions.

(i) For any given $V$, we have%
\begin{equation}
\left[ \eta ^{+},H\right] \neq 0,
\end{equation}%
and%
\begin{equation}
\left[ \eta ^{+},\left[ \eta ^{+},H\right] \right] \neq 0,
\end{equation}%
but 
\begin{equation}
\left[ \eta ^{+},\left[ \eta ^{+},\left[ \eta ^{+},H\right] \right] \right]
=0.
\end{equation}%
In addition, we have%
\begin{equation}
H\left\vert 0\right\rangle =0,
\end{equation}%
where $\left\vert 0\right\rangle $ is the vacuum state of the hardcore boson
operator. So far, these relations do not guarantee the construction of
eigenstates by the operator $\eta ^{+}$. To proceed, the following condition
must be satisfied.

(ii) In the large $V$ limit $\left( V\rightarrow \infty \right) $, applying
the above non-zero commutation relations to the vacuum state gives%
\begin{equation}
\left[ \eta ^{+},H\right] \left\vert 0\right\rangle =0,
\end{equation}%
and%
\begin{equation}
\left[ \eta ^{+},\left[ \eta ^{+},H\right] \right] \left\vert 0\right\rangle
=0.
\end{equation}%
We note that this system meets the conditions of the 2nd-order RSGA of \cite%
{moudgalya2020eta}. Therefore, a set of eigenstates can be constructed by
the operator $\eta ^{+}$. In the following, we would like to present the
conclusion more clearly. 
\begin{figure*}[t]
\centering
\includegraphics[width=1\textwidth]{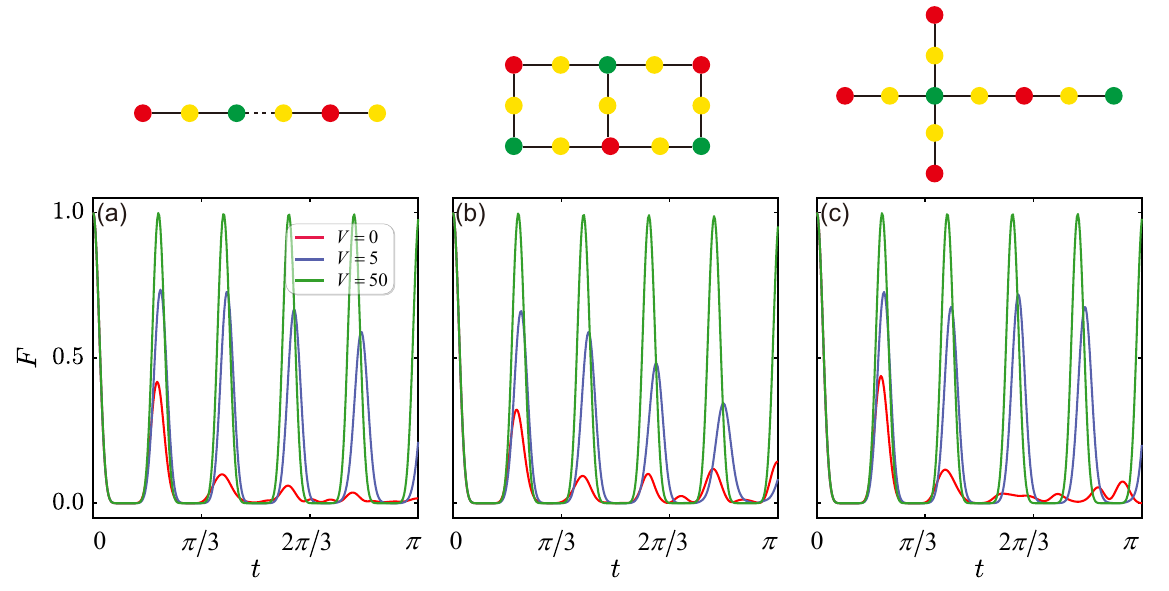}
\caption{Plots of the fidelity of dynamical evolution for Hamiltonians with
different structures (see Eq. (\protect\ref{H})) with the initial state
given by Eq. (\protect\ref{psi0}). The structures corresponding to Figs.
(a), (b), and (c) are labeled above each figure. (a), we take $N_{a}=3$, $%
N_{b}=3$, and $N_{c}=5$. (b), we take $N_{a}=3$, $N_{b}=3$, and $N_{c}=7$.
(c), we take $N_{a}=4$, $N_{b}=2$, and $N_{c}=5$. In all three figures, we
fix the hopping term $\protect\kappa=1$ and the chemical potential $\protect%
\mu=10$, and vary the interaction strength $V$ to plot the results. The
results show that different interaction strengths $V$ indeed have an impact
on the periodicity of the dynamics.}
\label{fig2}
\end{figure*}
We express our result in an alternative way. We propose a Hamiltonian on a
generalized Lieb lattice

\begin{equation}
\mathcal{H}=\sum_{i,j=1}^{N_{a},N_{b}}\kappa _{ij}\left( \alpha
_{i}^{\dagger }+\beta _{j}^{\dagger }\right) \gamma _{ij}+\text{\textrm{H.c.}%
},  \label{const H}
\end{equation}%
where $\alpha _{i}$, $\beta _{j}$, and $\gamma _{ij}$ are constrained
hardcore boson annihilation operators. The additional constraints on these
operators are the prohibition of nearest-neighbor pairs, i.e., 
\begin{equation}
\alpha _{i}\gamma _{ij}=\beta _{j}\gamma _{ij}=0.
\end{equation}%
In the above analysis, such constraints are applied by the infinite strength
of density-density interactions. Further discussion of the Hamiltonian $%
\mathcal{H}$ in a fermionic representation will be given in Sec. \ref%
{Connection to Fermi Hubbard model}. In this context, the introduction of
the operators $\{\alpha _{i},\beta _{j},\gamma _{ij}\}$ does not provide any
physical insight. It merely offers a concise presentation. Then we conclude
that a set of the eigenstates of $\mathcal{H}$ can be expressed in the form%
\begin{equation}
\left\vert \psi ^{m}\right\rangle =\frac{1}{m!\sqrt{C_{N_{a}+N_{b}}^{m}}}%
(\sum_{i=1}^{N_{a}}\alpha _{i}^{\dagger }-\sum_{j=1}^{N_{b}}\beta
_{j}^{\dagger })^{m}\left\vert 0\right\rangle ,
\end{equation}%
with $m\in \left[ 0,N_{a}+N_{b}\right] $. They are degenerate eigenstates
with zero energy. They are also the eigenstates of the particle number
operator $\sum_{i=1}^{N_{a}}\alpha _{i}^{\dagger }\alpha
_{i}-\sum_{j=1}^{N_{b}}\beta _{j}^{\dagger }\beta _{j}$. The set of
eigenstates $\left\{ \left\vert \psi ^{m}\right\rangle \right\} $
constitutes an invariant subspace, which is not based on the symmetry of the
system. When a uniform chemical potential is added, these degenerate
eigenstates form an energy tower.

\section{Weak Hilbert space fragmentation}

\label{Weak Hilbert space fragmentation}

The analysis in the last section indicates that the HSF is induced by the
prohibition of two nearest-neighbor pair configurations. This is achieved by
setting $V$ to be infinite in the Hamiltonian, as given by Eq. (\ref{H}).
One would presumably expect the HSF to become weak when finite $V$ is
taken. In this section, we investigate how the value of $V$ affects the
efficiency of the fragmentation. This analysis is performed from the
perspective of system dynamics. Specifically, we calculate the time
evolutions of a given initial state under various values of $V$ to
understand the dynamics of the system.

We consider the time evolution of an initial state given by

\begin{equation}
\left\vert \phi (0)\right\rangle
=2^{-(N_{a}+N_{b})/2}\prod_{i=1}^{N_{a}}(1+a_{i}^{\dagger
})\prod_{j=1}^{N_{b}}(1-b_{j}^{\dagger })\left\vert 0\right\rangle .
\label{psi0}
\end{equation}%
We choose this initial state for the following reasons. (i) Its time
evolution can be solved exactly under a special condition. (ii) The
corresponding dynamics for 1D system was studied experimentally \cite%
{jepsen2022long}. Indeed, the expression of $\left\vert \phi
(0)\right\rangle $ can be rewritten in the form%
\begin{equation}
\left\vert \phi (0)\right\rangle
=2^{-(N_{a}+N_{b})/2}\sum_{m=0}^{N_{a}+N_{b}}\sqrt{C_{N_{a}+N_{b}}^{m}}%
\left\vert \psi ^{m}\right\rangle .
\end{equation}%
The evolved state $\left\vert \phi (t)\right\rangle =e^{-iHt}\left\vert \phi
(0)\right\rangle $ under the Hamiltonian in the large $V$ limit, with
chemical potentials, given by

\begin{equation}
H\rightarrow H+\mu \sum_{i,j=1}^{N_{a},N_{b}}\left( a_{i}^{\dagger
}a_{i}+b_{j}^{\dagger }b_{j}+c_{ij}^{\dagger }c_{ij}\right) .
\end{equation}%
can be expressed in the form%
\begin{eqnarray}
&&\left\vert \phi (t)\right\rangle
=2^{-(N_{a}+N_{b})/2}\sum_{m=0}^{N_{a}+N_{b}}\sqrt{C_{N_{a}+N_{b}}^{m}}%
e^{-im\mu t}\left\vert \psi ^{m}\right\rangle  \notag \\
&=&2^{-(N_{a}+N_{b})/2}\prod_{i=1}^{N_{a}}(1+e^{-i\mu t}a_{i}^{\dagger
})\prod_{j=1}^{N_{b}}(1-e^{-i\mu t}b_{j}^{\dagger })\left\vert
0\right\rangle .
\end{eqnarray}%
We note that $\left\vert \phi (t)\right\rangle $ remains a simple product
state. It is periodic, indicating perfect Hilbert-space fragmentation.
Moreover, this phenomenon can be detected by measuring a single-site state.
We employ the fidelity, the squared modulus of the overlap between the two
states $\left\vert \phi (0)\right\rangle $ and $\left\vert \phi
(t)\right\rangle $,%
\begin{equation}
F(t)=\left\vert \langle \phi (t)\left\vert \phi (0)\right\rangle \right\vert
^{2},  \label{Fg}
\end{equation}%
to quantify fragmentation in the finite-$V$ case. To demonstrate this, we
perform numerical simulations on finite-size quantum spin Lieb lattices.
Several cluster types are considered, as illustrated in Fig. \ref{fig2}.
Each configuration has distinct values of $N_{a}$, $N_{b}$, and $N_{c}$.
Fig. \ref{fig2} also plots $F(t)$ for representative values of $V$. We draw
the following conclusions. (i) When $V$ is sufficiently large, $F(t)$
exhibits perfect periodic patterns for every configuration. (ii) For
intermediate $V$, quasi-periodic behavior of $F(t)$ emerges. (iii) In the $%
V=0$ limit, $F(t)$ loses all periodicity. These results show that, in
addition to the infinite-$V$ limit, weak Hilbert-space fragmentation also
appears at intermediate $V$, with the corresponding eigenstates forming
quasi-energy towers that act as quantum scars.

\section{Connection to Fermi Hubbard model}

\label{Connection to Fermi Hubbard model}

We know that the studied model can be mapped to the spin-1/2 XXZ model \cite%
{matsubara1956lattice}, which allows our results to be applied to both
hardcore boson and quantum spin systems. These correspond to itinerant
bosonic and localized fermionic systems. Specifically, these correspond to
itinerant bosonic and localized fermionic systems. In this section, we turn
to interacting fermionic systems and show that the Hamiltonian $\mathcal{H}$
given in Eq. (\ref{const H}) can serve as the effective Hamiltonian of a
Fermi Hubbard model in the weak-hopping limit.

We start with the fermion representation of the constrained hardcore boson
operators given in Eq. (\ref{const H}) to explore the underlying physics.
We introduce the transformation as 
\begin{eqnarray}
\alpha _{i} &=&c_{A,i,\uparrow }c_{A,i,\downarrow },  \notag \\
\beta _{j} &=&c_{B,j,\uparrow }c_{B,j,\downarrow },  \notag \\
\gamma _{ij} &=&c_{A,i,\uparrow }c_{B,j,\downarrow },
\label{f representation}
\end{eqnarray}%
where the operator $c_{\lambda ,j,\sigma }$ ($\lambda =A,B$) is the
annihilation operator of a spin-$\sigma $ fermion at site $j$, satisfying
the usual fermion anticommutation relations $\{c_{\lambda ,j,\sigma
}^{\dagger },$ $c_{\lambda ^{\prime },j^{\prime },\sigma ^{\prime
}}\}=\delta _{\lambda \lambda ^{\prime }}\delta _{jj^{\prime }}\delta
_{\sigma \sigma ^{\prime }}$ and $\{c_{\lambda ,j,\sigma },$ $c_{\lambda
^{\prime },j^{\prime },\sigma ^{\prime }}\}=0$. Intuitively, a simple way to
establish a fermionic version of the Hamiltonian $\mathcal{H}$ is to
directly replace the constrained hardcore boson operators with the fermion
operators via the above transformations. However, the obtained Hamiltonian
is somewhat challenging to realized in practice.

Indeed, substituting the transformations given by Eq. (\ref{f representation}%
) into the Hamiltonian $\mathcal{H}$ given in Eq. (\ref{const H}), we have%
\begin{equation}
\mathcal{H}=\sum_{i,j=1}^{N_{a},N_{b}}\kappa _{ij}(c_{A,i,\downarrow
}^{\dagger }c_{B,j,\downarrow }n_{A,i,\uparrow }+c_{B,j,\uparrow }^{\dag
}c_{A,i,\uparrow }n_{B,j,\downarrow })+\text{\textrm{H.c.}}.
\end{equation}%
The physical picture is clear: the Hamiltonian describes a conditional
hopping between sites of the two sublattices A and B. Specifically, only the
following hoppings are permitted 
\begin{equation*}
\left\vert \uparrow \right\rangle _{A}\left\vert \downarrow \right\rangle
_{B}\leftrightarrow \left\vert \uparrow \downarrow \right\rangle
_{A}\left\vert 0\right\rangle _{B}+\left\vert 0\right\rangle _{A}\left\vert
\uparrow \downarrow \right\rangle _{B},
\end{equation*}%
where states are given by $\left\vert \uparrow \right\rangle _{\lambda
}=c_{\lambda ,i,\uparrow }^{\dagger }\left\vert 0\right\rangle _{\lambda }$, 
$\left\vert \downarrow \right\rangle _{\lambda }=c_{\lambda ,i,\downarrow
}^{\dagger }\left\vert 0\right\rangle _{\lambda }$, and $\left\vert \uparrow
\downarrow \right\rangle _{\lambda }=c_{\lambda ,i,\uparrow }^{\dagger
}c_{\lambda ,i,\downarrow }^{\dagger }\left\vert 0\right\rangle _{\lambda }$%
. Here, the states $\left\vert 0\right\rangle _{A}$ and $\left\vert
0\right\rangle _{B}$ are vacuum states of the operators $c_{A,i,\sigma }$\
and $c_{B,j,\sigma }$, respectively.

The corresponding operator $\eta ^{+}$ becomes%
\begin{equation}
\eta ^{+}=\sum_{i=1}^{N_{a}}c_{A,i,\downarrow }^{\dag }c_{A,i,\uparrow
}^{\dag }-\sum_{j=1}^{N_{b}}c_{B,j,\downarrow }^{\dag }c_{B,j,\uparrow
}^{\dag },
\end{equation}%
based on which, we can verify that the Hamiltonian $\mathcal{H}$ possesses
the RSGA. Indeed, direct derivations show that
\begin{equation}
\left[ \eta ^{+},\mathcal{H}\right] \neq 0,
\end{equation}%
but%
\begin{equation}
\left[ \eta ^{+},\left[ \eta ^{+},\mathcal{H}\right] \right] =0,
\end{equation}%
and%
\begin{equation}
\left[ \eta ^{+},\mathcal{H}\right] \left\vert 0\right\rangle _{A}\left\vert
0\right\rangle _{B}=0.
\end{equation}%
\begin{figure}[th]
\centering
\includegraphics[width=0.51\textwidth]{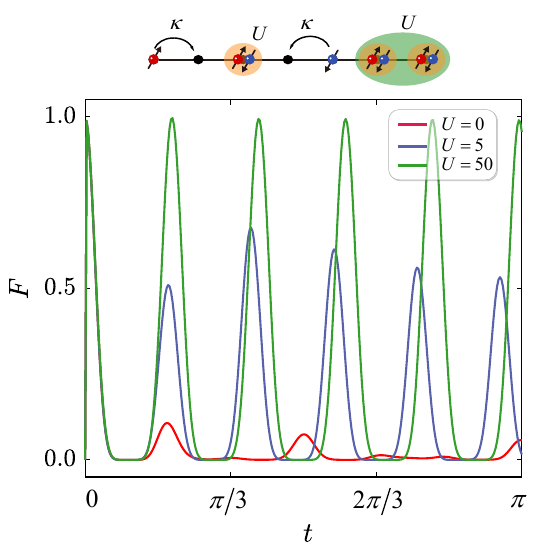}
\caption{Plots of the fidelity of the time evolution driven by the Fermi
Hubbard Hamiltonian, given by Eq. (\protect\ref{H_Hubbard}). The chain
system is illustrated at the top of the figure. The orange shading indicates
the on-site repulsion, and the green shading indicates the doublon-doublon
interaction. The initial state is $\left\vert \protect\phi (0)\right\rangle
=2^{-2}\sum_{m=0}^{4}\protect\sqrt{C_{4}^{m}}\left\vert \protect\psi %
^{m}\right\rangle $, where $\left\vert \protect\psi ^{m}\right\rangle $ is
given by Eq. (\protect\ref{FHM_IS}). The system parameters are $\protect\mu %
=5$, $N=7$, and $\protect\kappa =1$. Three representative values of $U$ are
indicated in the figure.}
\label{fig3}
\end{figure}

These relations allow us to propose a modified Fermi Hubbard model with the
Hamiltonian

\begin{equation}
H_{\text{\textrm{FH}}}=\mathcal{H}+U\sum_{l=1}^{N_{a}+N_{b}}\left(
n_{l,\uparrow }n_{l,\downarrow }-\frac{n_{l,\uparrow }+n_{l,\downarrow }}{2}%
\right)  \label{Hubbard AB}
\end{equation}%
where the parameter $\kappa $ denotes the hopping amplitude, $U$ is the
strength of both the on-site repulsion. It is an variant of the conventional
Hubbard model \cite{yang1989eta,yang1990so} by imposing the constraint on the hopping
terms, for instance, $c_{A,i,\downarrow }^{\dagger }c_{B,j,\downarrow
}\rightarrow c_{A,i,\downarrow }^{\dagger }c_{B,j,\downarrow
}n_{A,i,\uparrow }$. It is easy to check that $\left[ \eta ^{+},H_{\text{%
\textrm{FH}}}\right] \neq 0$ but $\left[ \eta ^{+},\left[ \eta ^{+},H_{\text{%
\textrm{FH}}}\right] \right] =0$ and $\left[ \eta ^{+},H_{\text{\textrm{FH}}}%
\right] \left\vert 0\right\rangle _{A}\left\vert 0\right\rangle _{B}=0$.
Obviously, the Hamiltonian $H_{\text{\textrm{FH}}}$ still meets the RSGA.
Then the set of $\eta $-pairing state%
\begin{equation}
\left( \sum_{i=1}^{N_{a}}c_{A,i,\downarrow }^{\dag }c_{A,i,\uparrow }^{\dag
}-\sum_{j=1}^{N_{b}}c_{B,j,\downarrow }^{\dag }c_{B,j,\uparrow }^{\dag
}\right) ^{m}\left\vert 0\right\rangle ,
\end{equation}%
are both eigenstates of the conventional Hubbard model and $H_{\text{\textrm{FH}}}$. Unlike
the situation in the conventional Hubbard model, this set of eigenstates for 
$H_{\text{\textrm{FH}}}$ is not supported by the $\eta $-pairing symmetry,
which has been extensively studied \cite{he2025eta,zhang2021topologically,wang2024flat,yang2022dynamic,zhang2021eta,zhang2020dynamical,moudgalya2020eta,li2020eta,kaneko2020charge,kaneko2019photoinduced}. This
example reveals a clear connection between the RSGA and HSF.

We would like to point out that the transformations given by Eq. (\ref{f
representation}) is not unique for the investigation of RSGA in Fermi
system. In the following, we propose another modified Hubbard model whose
effective Hamiltonian corresponds to $\mathcal{H}$ under certain conditions. 
{The main strategy is to construct a Hamiltonian that places the hard-core
bosons formed by each pair of fermions in the same energy shell. This
ensures that other types of configurations can be ruled out when considering
only a certain energy scale.} For simplicity, we consider a one-dimensional
Fermi Hubbard model to demonstrate our strategy. The conclusions we obtain
are applicable to a general system on a bipartite lattice.

We consider a Fermi Hubbard chain model with a doublon-doublon interaction,
whose Hamiltonian is%
\begin{eqnarray}
&&H_{\text{\textrm{FH}}}=\kappa \sum_{l=1,\sigma =\uparrow ,\downarrow
}^{N}(c_{l,\sigma }^{\dagger }c_{l+1,\sigma }+\text{\textrm{H.c.}}%
)+U\sum_{l=1}^{N}(n_{l,\uparrow }n_{l,\downarrow }  \notag \\
&&-\frac{n_{l,\uparrow }+n_{l,\downarrow }}{2})+U%
\sum_{l=1}^{N-1}n_{l}^{d}n_{l+1}^{d}+\mu \sum_{l=1,\sigma =\uparrow
,\downarrow }^{N}n_{l,\sigma },  \label{H_Hubbard}
\end{eqnarray}%
where the parameter $\kappa $ denotes the hopping amplitude, $U$ is the
strength of both the on-site repulsion and the nearest-neighbor (NN)
doublon-doublon interaction. Here, $n_{l}^{d}=d_{l}^{\dagger }d_{l}$ is
doublon number operator, where $d_{l}^{\dag }=c_{l,\downarrow }^{\dagger
}c_{l,\downarrow }^{\dagger }$ creates a double occupancy at site $l$. The
chain system is illustrated at the top of Fig. \ref{fig3}. Three
representative values of $U$ are indicated in the figure. In the following,
the size of the chain, $N$, is restricted to be odd. In contrast to the
conventional Hubbard model, there exists a doublon-doublon interaction. We
focus on the features of the Hamiltonian $H_{\text{\textrm{FH}}}$ in the
zero-energy regime under the condition $\kappa \ll U$ and $\mu =0$. To this
end, we consider the eigenstates of $H_{\text{\textrm{FH}}}$ with zero $%
\kappa $. We find that these states can be expressed in the form%
\begin{equation}
\prod_{\left\{ l\right\} }c_{l,\downarrow }^{\dag }c_{l,\uparrow }^{\dag
}\left\vert 0\right\rangle .  \label{subspace}
\end{equation}
For nonzero $\kappa $, the effective Hamiltonian is
\begin{equation}
H_{\text{eff}}=\sum_{j=1}^{N}\frac{4\kappa ^{2}}{U}d_{j}^{\dagger }d_{j+1}+%
\text{\textrm{H.c.}}+U\sum_{l=1}^{N-1}n_{l}^{d}n_{l+1}^{d},
\end{equation}%
which is obtained from the second-order perturbation method. It is
essentially the matrix representation of the Hamiltonian $H_{\text{\textrm{FH%
}}}$ in the subspace spanned by the states given by Eq. (\ref{subspace}).
This can also be regarded as a truncation approximation. In this context the
effective Hamiltonian can be written as $H_{\text{eff}}=PH_{\text{\textrm{FH}%
}}P^{-1}$, where the project operator $P$ projects onto the subspace
spanned by the states given by Eq. (\ref{subspace}).

The corresponding operator $\eta ^{+}$ becomes%
\begin{equation}
\eta ^{+}=\sum_{l=1}^{(N+1)/2}\left( -1\right) ^{l}c_{2l-1,\downarrow
}^{\dag }c_{2l-1,\uparrow }^{\dag },
\end{equation}%
based on which we can verify that the Hamiltonian $H_{\text{eff}}$ exhibits
the RSGA. Accordingly, the set of states 
\begin{equation}
\left\vert \psi ^{m}\right\rangle =\frac{1}{m!\sqrt{C_{(N+1)/2}^{m}}}\left(
\eta ^{+}\right) ^{m}\left\vert 0\right\rangle ,  \label{FHM_IS}
\end{equation}%
with $m\in \left[ 0,(N+1)/2\right] $, forms a degenerate set of eigenstates
of the Hamiltonian $H_{\text{eff}}$. Note that, $\left\{ \left\vert \psi
^{m}\right\rangle \right\} $ is an alternative set of $\eta $-pairing
states and are not eigenstate of $H_{\text{\textrm{FH}}}$. This indicates
that the Hilbert space is approximately fragmented by increasing $U$. In the
following, we investigate how the value of $U$ affects the efficiency of the
effective Hamiltonian $H_{\text{eff}}$, as well as the fragmentation. We
still perform this task by examining the time evolutions of a given initial
state under various values of $U$. The initial state is set to be the
superposition of $\left\vert \psi ^{m}\right\rangle $ given by Eq. (\ref%
{FHM_IS}). The driven systems are governed by $H_{\text{\textrm{FH}}}$, that
is, $\left\vert \phi (t)\right\rangle =e^{-iH_{\text{\textrm{FH}}%
}t}\left\vert \phi (0)\right\rangle $. Fig. \ref{fig3} shows the
corresponding plots $F(t)$ for representative values of $U$. We draw the
following conclusions. (i) When $U$ is sufficiently large, $F(t)$ exhibits a
perfect periodic pattern. (ii) For intermediate $U$, quasi-periodic behavior
of $F(t)$ emerges. (iii) In the $U=0$ case, $F(t)$ loses all periodicity.
These results show that, in addition to the infinite-$U$ limit, weak
Hilbert-space fragmentation also appears at intermediate $U$, with the
corresponding eigenstates forming quasi-energy towers that act as quantum
scars. These results also enrich the Fermi models, which possess $\eta $%
-pairing eigenstates without the need for $\eta $-pairing symmetry.

\section{Summary}

\label{Summary}
In summary, we have engaged with two types of models: the
hardcore bosonic and the fermionic Hubbard models. We have proposed a
formalism that establishes a connection between HSF and energy towers for a
set of hardcore bosonic systems on a generalized Lieb lattice. The HSF is
conducted through constraints on neighboring pairs, while the energy towers
arise from the RSGA. We have also studied the application of this formalism
to the fermionic Hubbard model. This was achieved through the fermionic
representations of hardcore bosons. We have proposed two types of modified
Hubbard models to demonstrate our method. The HSF is conducted through
constraints on neighboring hopping and on doublon-doublon pairs,
respectively. Numerical simulations accord with our predictions. Our work
provides an explicit relationship between models featuring
interaction-induced constraints and the construction of energy towers.

\section*{DATA AVAILABILITY}
The data that support the findings of this article are openly
available \cite{dataset}.

\section*{Acknowledgment}

We acknowledge the support of NSFC (Grants No. 12374461).
\bibliographystyle{apsrev4-2.bst}
\bibliography{ref.bib}

\begin{thebibliography}{67}%
\makeatletter
\providecommand \@ifxundefined [1]{%
 \@ifx{#1\undefined}
}%
\providecommand \@ifnum [1]{%
 \ifnum #1\expandafter \@firstoftwo
 \else \expandafter \@secondoftwo
 \fi
}%
\providecommand \@ifx [1]{%
 \ifx #1\expandafter \@firstoftwo
 \else \expandafter \@secondoftwo
 \fi
}%
\providecommand \natexlab [1]{#1}%
\providecommand \enquote  [1]{``#1''}%
\providecommand \bibnamefont  [1]{#1}%
\providecommand \bibfnamefont [1]{#1}%
\providecommand \citenamefont [1]{#1}%
\providecommand \href@noop [0]{\@secondoftwo}%
\providecommand \href [0]{\begingroup \@sanitize@url \@href}%
\providecommand \@href[1]{\@@startlink{#1}\@@href}%
\providecommand \@@href[1]{\endgroup#1\@@endlink}%
\providecommand \@sanitize@url [0]{\catcode `\\12\catcode `\$12\catcode
  `\&12\catcode `\#12\catcode `\^12\catcode `\_12\catcode `\%12\relax}%
\providecommand \@@startlink[1]{}%
\providecommand \@@endlink[0]{}%
\providecommand \url  [0]{\begingroup\@sanitize@url \@url }%
\providecommand \@url [1]{\endgroup\@href {#1}{\urlprefix }}%
\providecommand \urlprefix  [0]{URL }%
\providecommand \Eprint [0]{\href }%
\providecommand \doibase [0]{https://doi.org/}%
\providecommand \selectlanguage [0]{\@gobble}%
\providecommand \bibinfo  [0]{\@secondoftwo}%
\providecommand \bibfield  [0]{\@secondoftwo}%
\providecommand \translation [1]{[#1]}%
\providecommand \BibitemOpen [0]{}%
\providecommand \bibitemStop [0]{}%
\providecommand \bibitemNoStop [0]{.\EOS\space}%
\providecommand \EOS [0]{\spacefactor3000\relax}%
\providecommand \BibitemShut  [1]{\csname bibitem#1\endcsname}%
\let\auto@bib@innerbib\@empty
\bibitem [{\citenamefont {Deutsch}(1991)}]{Deutsch_Quantum}%
  \BibitemOpen
  \bibfield  {author} {\bibinfo {author} {\bibfnamefont {J.~M.}\ \bibnamefont
  {Deutsch}},\ }\href {https://doi.org/10.1103/PhysRevA.43.2046} {\bibfield
  {journal} {\bibinfo  {journal} {Phys. Rev. A}\ }\textbf {\bibinfo {volume}
  {43}},\ \bibinfo {pages} {2046} (\bibinfo {year} {1991})}\BibitemShut
  {NoStop}%
\bibitem [{\citenamefont {Srednicki}(1994)}]{srednicki1994chaos}%
  \BibitemOpen
  \bibfield  {author} {\bibinfo {author} {\bibfnamefont {M.}~\bibnamefont
  {Srednicki}},\ }\href@noop {} {\bibfield  {journal} {\bibinfo  {journal}
  {Phys. Rev. E}\ }\textbf {\bibinfo {volume} {50}},\ \bibinfo {pages} {888}
  (\bibinfo {year} {1994})}\BibitemShut {NoStop}%
\bibitem [{\citenamefont {D'Alessio}\ \emph {et~al.}(2016)\citenamefont
  {D'Alessio}, \citenamefont {Kafri}, \citenamefont {Polkovnikov},\ and\
  \citenamefont {Rigol}}]{d2016quantum}%
  \BibitemOpen
  \bibfield  {author} {\bibinfo {author} {\bibfnamefont {L.}~\bibnamefont
  {D'Alessio}}, \bibinfo {author} {\bibfnamefont {Y.}~\bibnamefont {Kafri}},
  \bibinfo {author} {\bibfnamefont {A.}~\bibnamefont {Polkovnikov}},\ and\
  \bibinfo {author} {\bibfnamefont {M.}~\bibnamefont {Rigol}},\ }\href@noop {}
  {\bibfield  {journal} {\bibinfo  {journal} {Advances in Physics}\ }\textbf
  {\bibinfo {volume} {65}},\ \bibinfo {pages} {239} (\bibinfo {year}
  {2016})}\BibitemShut {NoStop}%
\bibitem [{\citenamefont {Borgonovi}\ \emph {et~al.}(2016)\citenamefont
  {Borgonovi}, \citenamefont {Izrailev}, \citenamefont {Santos},\ and\
  \citenamefont {Zelevinsky}}]{borgonovi2016quantum}%
  \BibitemOpen
  \bibfield  {author} {\bibinfo {author} {\bibfnamefont {F.}~\bibnamefont
  {Borgonovi}}, \bibinfo {author} {\bibfnamefont {F.~M.}\ \bibnamefont
  {Izrailev}}, \bibinfo {author} {\bibfnamefont {L.~F.}\ \bibnamefont
  {Santos}},\ and\ \bibinfo {author} {\bibfnamefont {V.~G.}\ \bibnamefont
  {Zelevinsky}},\ }\href@noop {} {\bibfield  {journal} {\bibinfo  {journal}
  {Physics Reports}\ }\textbf {\bibinfo {volume} {626}},\ \bibinfo {pages} {1}
  (\bibinfo {year} {2016})}\BibitemShut {NoStop}%
\bibitem [{\citenamefont {Gogolin}\ and\ \citenamefont
  {Eisert}(2016)}]{gogolin2016equilibration}%
  \BibitemOpen
  \bibfield  {author} {\bibinfo {author} {\bibfnamefont {C.}~\bibnamefont
  {Gogolin}}\ and\ \bibinfo {author} {\bibfnamefont {J.}~\bibnamefont
  {Eisert}},\ }\href@noop {} {\bibfield  {journal} {\bibinfo  {journal}
  {Reports on Progress in Physics}\ }\textbf {\bibinfo {volume} {79}},\
  \bibinfo {pages} {056001} (\bibinfo {year} {2016})}\BibitemShut {NoStop}%
\bibitem [{\citenamefont {Serbyn}\ \emph {et~al.}(2021)\citenamefont {Serbyn},
  \citenamefont {Abanin},\ and\ \citenamefont {Papi{\'c}}}]{serbyn2021quantum}%
  \BibitemOpen
  \bibfield  {author} {\bibinfo {author} {\bibfnamefont {M.}~\bibnamefont
  {Serbyn}}, \bibinfo {author} {\bibfnamefont {D.~A.}\ \bibnamefont {Abanin}},\
  and\ \bibinfo {author} {\bibfnamefont {Z.}~\bibnamefont {Papi{\'c}}},\
  }\href@noop {} {\bibfield  {journal} {\bibinfo  {journal} {Nature Physics}\
  }\textbf {\bibinfo {volume} {17}},\ \bibinfo {pages} {675} (\bibinfo {year}
  {2021})}\BibitemShut {NoStop}%
\bibitem [{\citenamefont {Nandkishore}\ and\ \citenamefont
  {Huse}(2015)}]{nandkishore2015many}%
  \BibitemOpen
  \bibfield  {author} {\bibinfo {author} {\bibfnamefont {R.}~\bibnamefont
  {Nandkishore}}\ and\ \bibinfo {author} {\bibfnamefont {D.~A.}\ \bibnamefont
  {Huse}},\ }\href@noop {} {\bibfield  {journal} {\bibinfo  {journal} {Annu.
  Rev. Condens. Matter Phys.}\ }\textbf {\bibinfo {volume} {6}},\ \bibinfo
  {pages} {15} (\bibinfo {year} {2015})}\BibitemShut {NoStop}%
\bibitem [{\citenamefont {Abanin}\ \emph {et~al.}(2019)\citenamefont {Abanin},
  \citenamefont {Altman}, \citenamefont {Bloch},\ and\ \citenamefont
  {Serbyn}}]{abanin2019colloquium}%
  \BibitemOpen
  \bibfield  {author} {\bibinfo {author} {\bibfnamefont {D.~A.}\ \bibnamefont
  {Abanin}}, \bibinfo {author} {\bibfnamefont {E.}~\bibnamefont {Altman}},
  \bibinfo {author} {\bibfnamefont {I.}~\bibnamefont {Bloch}},\ and\ \bibinfo
  {author} {\bibfnamefont {M.}~\bibnamefont {Serbyn}},\ }\href@noop {}
  {\bibfield  {journal} {\bibinfo  {journal} {Reviews of Modern Physics}\
  }\textbf {\bibinfo {volume} {91}},\ \bibinfo {pages} {021001} (\bibinfo
  {year} {2019})}\BibitemShut {NoStop}%
\bibitem [{\citenamefont {Bernien}\ \emph {et~al.}(2017)\citenamefont
  {Bernien}, \citenamefont {Schwartz}, \citenamefont {Keesling}, \citenamefont
  {Levine}, \citenamefont {Omran}, \citenamefont {Pichler}, \citenamefont
  {Choi}, \citenamefont {Zibrov}, \citenamefont {Endres}, \citenamefont
  {Greiner} \emph {et~al.}}]{bernien2017probing}%
  \BibitemOpen
  \bibfield  {author} {\bibinfo {author} {\bibfnamefont {H.}~\bibnamefont
  {Bernien}}, \bibinfo {author} {\bibfnamefont {S.}~\bibnamefont {Schwartz}},
  \bibinfo {author} {\bibfnamefont {A.}~\bibnamefont {Keesling}}, \bibinfo
  {author} {\bibfnamefont {H.}~\bibnamefont {Levine}}, \bibinfo {author}
  {\bibfnamefont {A.}~\bibnamefont {Omran}}, \bibinfo {author} {\bibfnamefont
  {H.}~\bibnamefont {Pichler}}, \bibinfo {author} {\bibfnamefont
  {S.}~\bibnamefont {Choi}}, \bibinfo {author} {\bibfnamefont {A.~S.}\
  \bibnamefont {Zibrov}}, \bibinfo {author} {\bibfnamefont {M.}~\bibnamefont
  {Endres}}, \bibinfo {author} {\bibfnamefont {M.}~\bibnamefont {Greiner}},
  \emph {et~al.},\ }\href@noop {} {\bibfield  {journal} {\bibinfo  {journal}
  {Nature}\ }\textbf {\bibinfo {volume} {551}},\ \bibinfo {pages} {579}
  (\bibinfo {year} {2017})}\BibitemShut {NoStop}%
\bibitem [{\citenamefont {Choi}\ \emph {et~al.}(2019)\citenamefont {Choi},
  \citenamefont {Turner}, \citenamefont {Pichler}, \citenamefont {Ho},
  \citenamefont {Michailidis}, \citenamefont {Papi{\'c}}, \citenamefont
  {Serbyn}, \citenamefont {Lukin},\ and\ \citenamefont
  {Abanin}}]{choi2019emergent}%
  \BibitemOpen
  \bibfield  {author} {\bibinfo {author} {\bibfnamefont {S.}~\bibnamefont
  {Choi}}, \bibinfo {author} {\bibfnamefont {C.~J.}\ \bibnamefont {Turner}},
  \bibinfo {author} {\bibfnamefont {H.}~\bibnamefont {Pichler}}, \bibinfo
  {author} {\bibfnamefont {W.~W.}\ \bibnamefont {Ho}}, \bibinfo {author}
  {\bibfnamefont {A.~A.}\ \bibnamefont {Michailidis}}, \bibinfo {author}
  {\bibfnamefont {Z.}~\bibnamefont {Papi{\'c}}}, \bibinfo {author}
  {\bibfnamefont {M.}~\bibnamefont {Serbyn}}, \bibinfo {author} {\bibfnamefont
  {M.~D.}\ \bibnamefont {Lukin}},\ and\ \bibinfo {author} {\bibfnamefont
  {D.~A.}\ \bibnamefont {Abanin}},\ }\href@noop {} {\bibfield  {journal}
  {\bibinfo  {journal} {Phys. Rev. L}\ }\textbf {\bibinfo {volume} {122}},\
  \bibinfo {pages} {220603} (\bibinfo {year} {2019})}\BibitemShut {NoStop}%
\bibitem [{\citenamefont {Turner}\ \emph
  {et~al.}(2018{\natexlab{a}})\citenamefont {Turner}, \citenamefont
  {Michailidis}, \citenamefont {Abanin}, \citenamefont {Serbyn},\ and\
  \citenamefont {Papi{\'c}}}]{turner2018weak}%
  \BibitemOpen
  \bibfield  {author} {\bibinfo {author} {\bibfnamefont {C.~J.}\ \bibnamefont
  {Turner}}, \bibinfo {author} {\bibfnamefont {A.~A.}\ \bibnamefont
  {Michailidis}}, \bibinfo {author} {\bibfnamefont {D.~A.}\ \bibnamefont
  {Abanin}}, \bibinfo {author} {\bibfnamefont {M.}~\bibnamefont {Serbyn}},\
  and\ \bibinfo {author} {\bibfnamefont {Z.}~\bibnamefont {Papi{\'c}}},\
  }\href@noop {} {\bibfield  {journal} {\bibinfo  {journal} {Nature Physics}\
  }\textbf {\bibinfo {volume} {14}},\ \bibinfo {pages} {745} (\bibinfo {year}
  {2018}{\natexlab{a}})}\BibitemShut {NoStop}%
\bibitem [{\citenamefont {Pai}\ \emph {et~al.}(2019)\citenamefont {Pai},
  \citenamefont {Pretko},\ and\ \citenamefont
  {Nandkishore}}]{pai2019localization}%
  \BibitemOpen
  \bibfield  {author} {\bibinfo {author} {\bibfnamefont {S.}~\bibnamefont
  {Pai}}, \bibinfo {author} {\bibfnamefont {M.}~\bibnamefont {Pretko}},\ and\
  \bibinfo {author} {\bibfnamefont {R.~M.}\ \bibnamefont {Nandkishore}},\
  }\href@noop {} {\bibfield  {journal} {\bibinfo  {journal} {Phys. Rev. X}\
  }\textbf {\bibinfo {volume} {9}},\ \bibinfo {pages} {021003} (\bibinfo {year}
  {2019})}\BibitemShut {NoStop}%
\bibitem [{\citenamefont {Kwan}\ \emph {et~al.}(2025)\citenamefont {Kwan},
  \citenamefont {Wilhelm}, \citenamefont {Biswas},\ and\ \citenamefont
  {Parameswaran}}]{kwan2025minimal}%
  \BibitemOpen
  \bibfield  {author} {\bibinfo {author} {\bibfnamefont {Y.~H.}\ \bibnamefont
  {Kwan}}, \bibinfo {author} {\bibfnamefont {P.~H.}\ \bibnamefont {Wilhelm}},
  \bibinfo {author} {\bibfnamefont {S.}~\bibnamefont {Biswas}},\ and\ \bibinfo
  {author} {\bibfnamefont {S.}~\bibnamefont {Parameswaran}},\ }\href@noop {}
  {\bibfield  {journal} {\bibinfo  {journal} {Phys. Rev. L}\ }\textbf {\bibinfo
  {volume} {134}},\ \bibinfo {pages} {010411} (\bibinfo {year}
  {2025})}\BibitemShut {NoStop}%
\bibitem [{\citenamefont {Feng}\ and\ \citenamefont
  {Skinner}(2022)}]{feng2022hilbert}%
  \BibitemOpen
  \bibfield  {author} {\bibinfo {author} {\bibfnamefont {X.}~\bibnamefont
  {Feng}}\ and\ \bibinfo {author} {\bibfnamefont {B.}~\bibnamefont {Skinner}},\
  }\href@noop {} {\bibfield  {journal} {\bibinfo  {journal} {Phys. Rev. R}\
  }\textbf {\bibinfo {volume} {4}},\ \bibinfo {pages} {013053} (\bibinfo {year}
  {2022})}\BibitemShut {NoStop}%
\bibitem [{\citenamefont {Zhao}\ \emph {et~al.}(2020)\citenamefont {Zhao},
  \citenamefont {Vovrosh}, \citenamefont {Mintert},\ and\ \citenamefont
  {Knolle}}]{zhao2020quantum}%
  \BibitemOpen
  \bibfield  {author} {\bibinfo {author} {\bibfnamefont {H.}~\bibnamefont
  {Zhao}}, \bibinfo {author} {\bibfnamefont {J.}~\bibnamefont {Vovrosh}},
  \bibinfo {author} {\bibfnamefont {F.}~\bibnamefont {Mintert}},\ and\ \bibinfo
  {author} {\bibfnamefont {J.}~\bibnamefont {Knolle}},\ }\href@noop {}
  {\bibfield  {journal} {\bibinfo  {journal} {Phys. Rev. L}\ }\textbf {\bibinfo
  {volume} {124}},\ \bibinfo {pages} {160604} (\bibinfo {year}
  {2020})}\BibitemShut {NoStop}%
\bibitem [{\citenamefont {Turner}\ \emph {et~al.}(2021)\citenamefont {Turner},
  \citenamefont {Desaules}, \citenamefont {Bull},\ and\ \citenamefont
  {Papi{\'c}}}]{turner2021correspondence}%
  \BibitemOpen
  \bibfield  {author} {\bibinfo {author} {\bibfnamefont {C.~J.}\ \bibnamefont
  {Turner}}, \bibinfo {author} {\bibfnamefont {J.-Y.}\ \bibnamefont
  {Desaules}}, \bibinfo {author} {\bibfnamefont {K.}~\bibnamefont {Bull}},\
  and\ \bibinfo {author} {\bibfnamefont {Z.}~\bibnamefont {Papi{\'c}}},\
  }\href@noop {} {\bibfield  {journal} {\bibinfo  {journal} {Phys. Rev. X}\
  }\textbf {\bibinfo {volume} {11}},\ \bibinfo {pages} {021021} (\bibinfo
  {year} {2021})}\BibitemShut {NoStop}%
\bibitem [{\citenamefont {Mukherjee}\ \emph {et~al.}(2020)\citenamefont
  {Mukherjee}, \citenamefont {Nandy}, \citenamefont {Sen}, \citenamefont
  {Sen},\ and\ \citenamefont {Sengupta}}]{mukherjee2020collapse}%
  \BibitemOpen
  \bibfield  {author} {\bibinfo {author} {\bibfnamefont {B.}~\bibnamefont
  {Mukherjee}}, \bibinfo {author} {\bibfnamefont {S.}~\bibnamefont {Nandy}},
  \bibinfo {author} {\bibfnamefont {A.}~\bibnamefont {Sen}}, \bibinfo {author}
  {\bibfnamefont {D.}~\bibnamefont {Sen}},\ and\ \bibinfo {author}
  {\bibfnamefont {K.}~\bibnamefont {Sengupta}},\ }\href@noop {} {\bibfield
  {journal} {\bibinfo  {journal} {Phys. Rev. B}\ }\textbf {\bibinfo {volume}
  {101}},\ \bibinfo {pages} {245107} (\bibinfo {year} {2020})}\BibitemShut
  {NoStop}%
\bibitem [{\citenamefont {van Voorden}\ \emph {et~al.}(2020)\citenamefont {van
  Voorden}, \citenamefont {Min{\'a}{\v{r}}},\ and\ \citenamefont
  {Schoutens}}]{van2020quantum}%
  \BibitemOpen
  \bibfield  {author} {\bibinfo {author} {\bibfnamefont {B.}~\bibnamefont {van
  Voorden}}, \bibinfo {author} {\bibfnamefont {J.}~\bibnamefont
  {Min{\'a}{\v{r}}}},\ and\ \bibinfo {author} {\bibfnamefont {K.}~\bibnamefont
  {Schoutens}},\ }\href@noop {} {\bibfield  {journal} {\bibinfo  {journal}
  {Phys. Rev. B}\ }\textbf {\bibinfo {volume} {101}},\ \bibinfo {pages}
  {220305} (\bibinfo {year} {2020})}\BibitemShut {NoStop}%
\bibitem [{\citenamefont {Bluvstein}\ \emph {et~al.}(2021)\citenamefont
  {Bluvstein}, \citenamefont {Omran}, \citenamefont {Levine}, \citenamefont
  {Keesling}, \citenamefont {Semeghini}, \citenamefont {Ebadi}, \citenamefont
  {Wang}, \citenamefont {Michailidis}, \citenamefont {Maskara}, \citenamefont
  {Ho} \emph {et~al.}}]{bluvstein2021controlling}%
  \BibitemOpen
  \bibfield  {author} {\bibinfo {author} {\bibfnamefont {D.}~\bibnamefont
  {Bluvstein}}, \bibinfo {author} {\bibfnamefont {A.}~\bibnamefont {Omran}},
  \bibinfo {author} {\bibfnamefont {H.}~\bibnamefont {Levine}}, \bibinfo
  {author} {\bibfnamefont {A.}~\bibnamefont {Keesling}}, \bibinfo {author}
  {\bibfnamefont {G.}~\bibnamefont {Semeghini}}, \bibinfo {author}
  {\bibfnamefont {S.}~\bibnamefont {Ebadi}}, \bibinfo {author} {\bibfnamefont
  {T.~T.}\ \bibnamefont {Wang}}, \bibinfo {author} {\bibfnamefont {A.~A.}\
  \bibnamefont {Michailidis}}, \bibinfo {author} {\bibfnamefont
  {N.}~\bibnamefont {Maskara}}, \bibinfo {author} {\bibfnamefont {W.~W.}\
  \bibnamefont {Ho}}, \emph {et~al.},\ }\href@noop {} {\bibfield  {journal}
  {\bibinfo  {journal} {Science}\ }\textbf {\bibinfo {volume} {371}},\ \bibinfo
  {pages} {1355} (\bibinfo {year} {2021})}\BibitemShut {NoStop}%
\bibitem [{\citenamefont {Surace}\ \emph {et~al.}(2021)\citenamefont {Surace},
  \citenamefont {Votto}, \citenamefont {Lazo}, \citenamefont {Silva},
  \citenamefont {Dalmonte},\ and\ \citenamefont {Giudici}}]{surace2021exact}%
  \BibitemOpen
  \bibfield  {author} {\bibinfo {author} {\bibfnamefont {F.~M.}\ \bibnamefont
  {Surace}}, \bibinfo {author} {\bibfnamefont {M.}~\bibnamefont {Votto}},
  \bibinfo {author} {\bibfnamefont {E.~G.}\ \bibnamefont {Lazo}}, \bibinfo
  {author} {\bibfnamefont {A.}~\bibnamefont {Silva}}, \bibinfo {author}
  {\bibfnamefont {M.}~\bibnamefont {Dalmonte}},\ and\ \bibinfo {author}
  {\bibfnamefont {G.}~\bibnamefont {Giudici}},\ }\href@noop {} {\bibfield
  {journal} {\bibinfo  {journal} {Phys. Rev. B}\ }\textbf {\bibinfo {volume}
  {103}},\ \bibinfo {pages} {104302} (\bibinfo {year} {2021})}\BibitemShut
  {NoStop}%
\bibitem [{\citenamefont {Yang}\ \emph {et~al.}(2025)\citenamefont {Yang},
  \citenamefont {Magoni},\ and\ \citenamefont
  {Pichler}}]{yang2025constructing}%
  \BibitemOpen
  \bibfield  {author} {\bibinfo {author} {\bibfnamefont {F.}~\bibnamefont
  {Yang}}, \bibinfo {author} {\bibfnamefont {M.}~\bibnamefont {Magoni}},\ and\
  \bibinfo {author} {\bibfnamefont {H.}~\bibnamefont {Pichler}},\ }\href@noop
  {} {\bibfield  {journal} {\bibinfo  {journal} {arXiv preprint
  arXiv:2506.10806}\ } (\bibinfo {year} {2025})}\BibitemShut {NoStop}%
\bibitem [{\citenamefont {Schecter}\ and\ \citenamefont
  {Iadecola}(2019)}]{schecter2019weak}%
  \BibitemOpen
  \bibfield  {author} {\bibinfo {author} {\bibfnamefont {M.}~\bibnamefont
  {Schecter}}\ and\ \bibinfo {author} {\bibfnamefont {T.}~\bibnamefont
  {Iadecola}},\ }\href@noop {} {\bibfield  {journal} {\bibinfo  {journal}
  {Phys. Rev. L}\ }\textbf {\bibinfo {volume} {123}},\ \bibinfo {pages}
  {147201} (\bibinfo {year} {2019})}\BibitemShut {NoStop}%
\bibitem [{\citenamefont {Yang}\ \emph {et~al.}(2020)\citenamefont {Yang},
  \citenamefont {Liu}, \citenamefont {Gorshkov},\ and\ \citenamefont
  {Iadecola}}]{yang2020hilbert}%
  \BibitemOpen
  \bibfield  {author} {\bibinfo {author} {\bibfnamefont {Z.-C.}\ \bibnamefont
  {Yang}}, \bibinfo {author} {\bibfnamefont {F.}~\bibnamefont {Liu}}, \bibinfo
  {author} {\bibfnamefont {A.~V.}\ \bibnamefont {Gorshkov}},\ and\ \bibinfo
  {author} {\bibfnamefont {T.}~\bibnamefont {Iadecola}},\ }\href@noop {}
  {\bibfield  {journal} {\bibinfo  {journal} {Phys. Rev. L}\ }\textbf {\bibinfo
  {volume} {124}},\ \bibinfo {pages} {207602} (\bibinfo {year}
  {2020})}\BibitemShut {NoStop}%
\bibitem [{\citenamefont {Moudgalya}\ and\ \citenamefont
  {Motrunich}(2022)}]{moudgalya2022hilbert}%
  \BibitemOpen
  \bibfield  {author} {\bibinfo {author} {\bibfnamefont {S.}~\bibnamefont
  {Moudgalya}}\ and\ \bibinfo {author} {\bibfnamefont {O.~I.}\ \bibnamefont
  {Motrunich}},\ }\href@noop {} {\bibfield  {journal} {\bibinfo  {journal}
  {Phys. Rev. X}\ }\textbf {\bibinfo {volume} {12}},\ \bibinfo {pages} {011050}
  (\bibinfo {year} {2022})}\BibitemShut {NoStop}%
\bibitem [{\citenamefont {Li}\ \emph {et~al.}(2023)\citenamefont {Li},
  \citenamefont {Sala},\ and\ \citenamefont {Pollmann}}]{li2023hilbert}%
  \BibitemOpen
  \bibfield  {author} {\bibinfo {author} {\bibfnamefont {Y.}~\bibnamefont
  {Li}}, \bibinfo {author} {\bibfnamefont {P.}~\bibnamefont {Sala}},\ and\
  \bibinfo {author} {\bibfnamefont {F.}~\bibnamefont {Pollmann}},\ }\href@noop
  {} {\bibfield  {journal} {\bibinfo  {journal} {Phys. Rev. R}\ }\textbf
  {\bibinfo {volume} {5}},\ \bibinfo {pages} {043239} (\bibinfo {year}
  {2023})}\BibitemShut {NoStop}%
\bibitem [{\citenamefont {Francica}\ and\ \citenamefont
  {Dell'Anna}(2023)}]{francica2023hilbert}%
  \BibitemOpen
  \bibfield  {author} {\bibinfo {author} {\bibfnamefont {G.}~\bibnamefont
  {Francica}}\ and\ \bibinfo {author} {\bibfnamefont {L.}~\bibnamefont
  {Dell'Anna}},\ }\href@noop {} {\bibfield  {journal} {\bibinfo  {journal}
  {Phys. Rev. B}\ }\textbf {\bibinfo {volume} {108}},\ \bibinfo {pages}
  {045127} (\bibinfo {year} {2023})}\BibitemShut {NoStop}%
\bibitem [{\citenamefont {Nicolau}\ \emph {et~al.}(2023)\citenamefont
  {Nicolau}, \citenamefont {Marques}, \citenamefont {Dias},\ and\ \citenamefont
  {Ahufinger}}]{nicolau2023flat}%
  \BibitemOpen
  \bibfield  {author} {\bibinfo {author} {\bibfnamefont {E.}~\bibnamefont
  {Nicolau}}, \bibinfo {author} {\bibfnamefont {A.~M.}\ \bibnamefont
  {Marques}}, \bibinfo {author} {\bibfnamefont {R.~G.}\ \bibnamefont {Dias}},\
  and\ \bibinfo {author} {\bibfnamefont {V.}~\bibnamefont {Ahufinger}},\
  }\href@noop {} {\bibfield  {journal} {\bibinfo  {journal} {Phys. Rev. B}\
  }\textbf {\bibinfo {volume} {108}},\ \bibinfo {pages} {205104} (\bibinfo
  {year} {2023})}\BibitemShut {NoStop}%
\bibitem [{\citenamefont {Lesanovsky}\ and\ \citenamefont
  {Katsura}(2012)}]{lesanovsky2012interacting}%
  \BibitemOpen
  \bibfield  {author} {\bibinfo {author} {\bibfnamefont {I.}~\bibnamefont
  {Lesanovsky}}\ and\ \bibinfo {author} {\bibfnamefont {H.}~\bibnamefont
  {Katsura}},\ }\href@noop {} {\bibfield  {journal} {\bibinfo  {journal} {Phys.
  Rev. A}\ }\textbf {\bibinfo {volume} {86}},\ \bibinfo {pages} {041601}
  (\bibinfo {year} {2012})}\BibitemShut {NoStop}%
\bibitem [{\citenamefont {Lingenfelter}\ \emph {et~al.}(2024)\citenamefont
  {Lingenfelter}, \citenamefont {Yao}, \citenamefont {Pocklington},
  \citenamefont {Wang}, \citenamefont {Irfan}, \citenamefont {Pfaff},\ and\
  \citenamefont {Clerk}}]{lingenfelter2024exact}%
  \BibitemOpen
  \bibfield  {author} {\bibinfo {author} {\bibfnamefont {A.}~\bibnamefont
  {Lingenfelter}}, \bibinfo {author} {\bibfnamefont {M.}~\bibnamefont {Yao}},
  \bibinfo {author} {\bibfnamefont {A.}~\bibnamefont {Pocklington}}, \bibinfo
  {author} {\bibfnamefont {Y.-X.}\ \bibnamefont {Wang}}, \bibinfo {author}
  {\bibfnamefont {A.}~\bibnamefont {Irfan}}, \bibinfo {author} {\bibfnamefont
  {W.}~\bibnamefont {Pfaff}},\ and\ \bibinfo {author} {\bibfnamefont {A.~A.}\
  \bibnamefont {Clerk}},\ }\href@noop {} {\bibfield  {journal} {\bibinfo
  {journal} {Phys. Rev. X}\ }\textbf {\bibinfo {volume} {14}},\ \bibinfo
  {pages} {021028} (\bibinfo {year} {2024})}\BibitemShut {NoStop}%
\bibitem [{\citenamefont {Sala}\ \emph {et~al.}(2020)\citenamefont {Sala},
  \citenamefont {Rakovszky}, \citenamefont {Verresen}, \citenamefont {Knap},\
  and\ \citenamefont {Pollmann}}]{sala2020ergodicity}%
  \BibitemOpen
  \bibfield  {author} {\bibinfo {author} {\bibfnamefont {P.}~\bibnamefont
  {Sala}}, \bibinfo {author} {\bibfnamefont {T.}~\bibnamefont {Rakovszky}},
  \bibinfo {author} {\bibfnamefont {R.}~\bibnamefont {Verresen}}, \bibinfo
  {author} {\bibfnamefont {M.}~\bibnamefont {Knap}},\ and\ \bibinfo {author}
  {\bibfnamefont {F.}~\bibnamefont {Pollmann}},\ }\href@noop {} {\bibfield
  {journal} {\bibinfo  {journal} {Phys. Rev. X}\ }\textbf {\bibinfo {volume}
  {10}},\ \bibinfo {pages} {011047} (\bibinfo {year} {2020})}\BibitemShut
  {NoStop}%
\bibitem [{\citenamefont {Shiraishi}\ and\ \citenamefont
  {Mori}(2017)}]{shiraishi2017}%
  \BibitemOpen
  \bibfield  {author} {\bibinfo {author} {\bibfnamefont {N.}~\bibnamefont
  {Shiraishi}}\ and\ \bibinfo {author} {\bibfnamefont {T.}~\bibnamefont
  {Mori}},\ }\href@noop {} {\bibfield  {journal} {\bibinfo  {journal} {Physical
  review letters}\ }\textbf {\bibinfo {volume} {119}},\ \bibinfo {pages}
  {030601} (\bibinfo {year} {2017})}\BibitemShut {NoStop}%
\bibitem [{\citenamefont {Moudgalya}\ \emph
  {et~al.}(2018{\natexlab{a}})\citenamefont {Moudgalya}, \citenamefont
  {Rachel}, \citenamefont {Bernevig},\ and\ \citenamefont
  {Regnault}}]{moudgalya2018}%
  \BibitemOpen
  \bibfield  {author} {\bibinfo {author} {\bibfnamefont {S.}~\bibnamefont
  {Moudgalya}}, \bibinfo {author} {\bibfnamefont {S.}~\bibnamefont {Rachel}},
  \bibinfo {author} {\bibfnamefont {B.~A.}\ \bibnamefont {Bernevig}},\ and\
  \bibinfo {author} {\bibfnamefont {N.}~\bibnamefont {Regnault}},\ }\href@noop
  {} {\bibfield  {journal} {\bibinfo  {journal} {Physical Review B}\ }\textbf
  {\bibinfo {volume} {98}},\ \bibinfo {pages} {235155} (\bibinfo {year}
  {2018}{\natexlab{a}})}\BibitemShut {NoStop}%
\bibitem [{\citenamefont {Moudgalya}\ \emph
  {et~al.}(2018{\natexlab{b}})\citenamefont {Moudgalya}, \citenamefont
  {Regnault},\ and\ \citenamefont {Bernevig}}]{moudgalya20182}%
  \BibitemOpen
  \bibfield  {author} {\bibinfo {author} {\bibfnamefont {S.}~\bibnamefont
  {Moudgalya}}, \bibinfo {author} {\bibfnamefont {N.}~\bibnamefont
  {Regnault}},\ and\ \bibinfo {author} {\bibfnamefont {B.~A.}\ \bibnamefont
  {Bernevig}},\ }\href@noop {} {\bibfield  {journal} {\bibinfo  {journal}
  {Physical Review B}\ }\textbf {\bibinfo {volume} {98}},\ \bibinfo {pages}
  {235156} (\bibinfo {year} {2018}{\natexlab{b}})}\BibitemShut {NoStop}%
\bibitem [{\citenamefont {Khemani}\ \emph {et~al.}(2019)\citenamefont
  {Khemani}, \citenamefont {Laumann},\ and\ \citenamefont
  {Chandran}}]{khemani2019}%
  \BibitemOpen
  \bibfield  {author} {\bibinfo {author} {\bibfnamefont {V.}~\bibnamefont
  {Khemani}}, \bibinfo {author} {\bibfnamefont {C.~R.}\ \bibnamefont
  {Laumann}},\ and\ \bibinfo {author} {\bibfnamefont {A.}~\bibnamefont
  {Chandran}},\ }\href@noop {} {\bibfield  {journal} {\bibinfo  {journal}
  {Physical Review B}\ }\textbf {\bibinfo {volume} {99}},\ \bibinfo {pages}
  {161101} (\bibinfo {year} {2019})}\BibitemShut {NoStop}%
\bibitem [{\citenamefont {Ho}\ \emph {et~al.}(2019)\citenamefont {Ho},
  \citenamefont {Choi}, \citenamefont {Pichler},\ and\ \citenamefont
  {Lukin}}]{ho2019}%
  \BibitemOpen
  \bibfield  {author} {\bibinfo {author} {\bibfnamefont {W.~W.}\ \bibnamefont
  {Ho}}, \bibinfo {author} {\bibfnamefont {S.}~\bibnamefont {Choi}}, \bibinfo
  {author} {\bibfnamefont {H.}~\bibnamefont {Pichler}},\ and\ \bibinfo {author}
  {\bibfnamefont {M.~D.}\ \bibnamefont {Lukin}},\ }\href@noop {} {\bibfield
  {journal} {\bibinfo  {journal} {Physical review letters}\ }\textbf {\bibinfo
  {volume} {122}},\ \bibinfo {pages} {040603} (\bibinfo {year}
  {2019})}\BibitemShut {NoStop}%
\bibitem [{\citenamefont {Shibata}\ \emph {et~al.}(2020)\citenamefont
  {Shibata}, \citenamefont {Yoshioka},\ and\ \citenamefont
  {Katsura}}]{shibata2020}%
  \BibitemOpen
  \bibfield  {author} {\bibinfo {author} {\bibfnamefont {N.}~\bibnamefont
  {Shibata}}, \bibinfo {author} {\bibfnamefont {N.}~\bibnamefont {Yoshioka}},\
  and\ \bibinfo {author} {\bibfnamefont {H.}~\bibnamefont {Katsura}},\
  }\href@noop {} {\bibfield  {journal} {\bibinfo  {journal} {Physical Review
  Letters}\ }\textbf {\bibinfo {volume} {124}},\ \bibinfo {pages} {180604}
  (\bibinfo {year} {2020})}\BibitemShut {NoStop}%
\bibitem [{\citenamefont {McClarty}\ \emph {et~al.}(2020)\citenamefont
  {McClarty}, \citenamefont {Haque}, \citenamefont {Sen},\ and\ \citenamefont
  {Richter}}]{mcclarty2020}%
  \BibitemOpen
  \bibfield  {author} {\bibinfo {author} {\bibfnamefont {P.~A.}\ \bibnamefont
  {McClarty}}, \bibinfo {author} {\bibfnamefont {M.}~\bibnamefont {Haque}},
  \bibinfo {author} {\bibfnamefont {A.}~\bibnamefont {Sen}},\ and\ \bibinfo
  {author} {\bibfnamefont {J.}~\bibnamefont {Richter}},\ }\href@noop {}
  {\bibfield  {journal} {\bibinfo  {journal} {Physical Review B}\ }\textbf
  {\bibinfo {volume} {102}},\ \bibinfo {pages} {224303} (\bibinfo {year}
  {2020})}\BibitemShut {NoStop}%
\bibitem [{\citenamefont {Richter}\ and\ \citenamefont
  {Pal}(2022)}]{richter2022}%
  \BibitemOpen
  \bibfield  {author} {\bibinfo {author} {\bibfnamefont {J.}~\bibnamefont
  {Richter}}\ and\ \bibinfo {author} {\bibfnamefont {A.}~\bibnamefont {Pal}},\
  }\href@noop {} {\bibfield  {journal} {\bibinfo  {journal} {Physical Review
  Research}\ }\textbf {\bibinfo {volume} {4}},\ \bibinfo {pages} {L012003}
  (\bibinfo {year} {2022})}\BibitemShut {NoStop}%
\bibitem [{\citenamefont {Jeyaretnam}\ \emph {et~al.}(2021)\citenamefont
  {Jeyaretnam}, \citenamefont {Richter},\ and\ \citenamefont
  {Pal}}]{jeyaretnam2021}%
  \BibitemOpen
  \bibfield  {author} {\bibinfo {author} {\bibfnamefont {J.}~\bibnamefont
  {Jeyaretnam}}, \bibinfo {author} {\bibfnamefont {J.}~\bibnamefont
  {Richter}},\ and\ \bibinfo {author} {\bibfnamefont {A.}~\bibnamefont {Pal}},\
  }\href@noop {} {\bibfield  {journal} {\bibinfo  {journal} {Physical Review
  B}\ }\textbf {\bibinfo {volume} {104}},\ \bibinfo {pages} {014424} (\bibinfo
  {year} {2021})}\BibitemShut {NoStop}%
\bibitem [{\citenamefont {Turner}\ \emph
  {et~al.}(2018{\natexlab{b}})\citenamefont {Turner}, \citenamefont
  {Michailidis}, \citenamefont {Abanin}, \citenamefont {Serbyn},\ and\
  \citenamefont {Papi{\'c}}}]{turner20182}%
  \BibitemOpen
  \bibfield  {author} {\bibinfo {author} {\bibfnamefont {C.}~\bibnamefont
  {Turner}}, \bibinfo {author} {\bibfnamefont {A.}~\bibnamefont {Michailidis}},
  \bibinfo {author} {\bibfnamefont {D.}~\bibnamefont {Abanin}}, \bibinfo
  {author} {\bibfnamefont {M.}~\bibnamefont {Serbyn}},\ and\ \bibinfo {author}
  {\bibfnamefont {Z.}~\bibnamefont {Papi{\'c}}},\ }\href@noop {} {\bibfield
  {journal} {\bibinfo  {journal} {Physical Review B}\ }\textbf {\bibinfo
  {volume} {98}},\ \bibinfo {pages} {155134} (\bibinfo {year}
  {2018}{\natexlab{b}})}\BibitemShut {NoStop}%
\bibitem [{\citenamefont {Shiraishi}(2019)}]{shiraishi2019}%
  \BibitemOpen
  \bibfield  {author} {\bibinfo {author} {\bibfnamefont {N.}~\bibnamefont
  {Shiraishi}},\ }\href@noop {} {\bibfield  {journal} {\bibinfo  {journal}
  {Journal of Statistical Mechanics: Theory and Experiment}\ }\textbf {\bibinfo
  {volume} {2019}},\ \bibinfo {pages} {083103} (\bibinfo {year}
  {2019})}\BibitemShut {NoStop}%
\bibitem [{\citenamefont {Lin}\ and\ \citenamefont
  {Motrunich}(2019)}]{lin2019}%
  \BibitemOpen
  \bibfield  {author} {\bibinfo {author} {\bibfnamefont {C.-J.}\ \bibnamefont
  {Lin}}\ and\ \bibinfo {author} {\bibfnamefont {O.~I.}\ \bibnamefont
  {Motrunich}},\ }\href@noop {} {\bibfield  {journal} {\bibinfo  {journal}
  {Physical review letters}\ }\textbf {\bibinfo {volume} {122}},\ \bibinfo
  {pages} {173401} (\bibinfo {year} {2019})}\BibitemShut {NoStop}%
\bibitem [{\citenamefont {Khemani}\ \emph {et~al.}(2020)\citenamefont
  {Khemani}, \citenamefont {Hermele},\ and\ \citenamefont
  {Nandkishore}}]{khemani2020}%
  \BibitemOpen
  \bibfield  {author} {\bibinfo {author} {\bibfnamefont {V.}~\bibnamefont
  {Khemani}}, \bibinfo {author} {\bibfnamefont {M.}~\bibnamefont {Hermele}},\
  and\ \bibinfo {author} {\bibfnamefont {R.}~\bibnamefont {Nandkishore}},\
  }\href@noop {} {\bibfield  {journal} {\bibinfo  {journal} {Physical Review
  B}\ }\textbf {\bibinfo {volume} {101}},\ \bibinfo {pages} {174204} (\bibinfo
  {year} {2020})}\BibitemShut {NoStop}%
\bibitem [{\citenamefont {Dooley}\ and\ \citenamefont
  {Kells}(2020)}]{dooley2020}%
  \BibitemOpen
  \bibfield  {author} {\bibinfo {author} {\bibfnamefont {S.}~\bibnamefont
  {Dooley}}\ and\ \bibinfo {author} {\bibfnamefont {G.}~\bibnamefont {Kells}},\
  }\href@noop {} {\bibfield  {journal} {\bibinfo  {journal} {Physical Review
  B}\ }\textbf {\bibinfo {volume} {102}},\ \bibinfo {pages} {195114} (\bibinfo
  {year} {2020})}\BibitemShut {NoStop}%
\bibitem [{\citenamefont {Dooley}(2021)}]{dooley2021}%
  \BibitemOpen
  \bibfield  {author} {\bibinfo {author} {\bibfnamefont {S.}~\bibnamefont
  {Dooley}},\ }\href@noop {} {\bibfield  {journal} {\bibinfo  {journal} {PRX
  Quantum}\ }\textbf {\bibinfo {volume} {2}},\ \bibinfo {pages} {020330}
  (\bibinfo {year} {2021})}\BibitemShut {NoStop}%
\bibitem [{\citenamefont {Mark}\ \emph {et~al.}(2020)\citenamefont {Mark},
  \citenamefont {Lin},\ and\ \citenamefont {Motrunich}}]{mark2020unified}%
  \BibitemOpen
  \bibfield  {author} {\bibinfo {author} {\bibfnamefont {D.~K.}\ \bibnamefont
  {Mark}}, \bibinfo {author} {\bibfnamefont {C.-J.}\ \bibnamefont {Lin}},\ and\
  \bibinfo {author} {\bibfnamefont {O.~I.}\ \bibnamefont {Motrunich}},\
  }\href@noop {} {\bibfield  {journal} {\bibinfo  {journal} {Physical Review
  B}\ }\textbf {\bibinfo {volume} {101}},\ \bibinfo {pages} {195131} (\bibinfo
  {year} {2020})}\BibitemShut {NoStop}%
\bibitem [{\citenamefont {Moudgalya}\ \emph {et~al.}(2020)\citenamefont
  {Moudgalya}, \citenamefont {Regnault},\ and\ \citenamefont
  {Bernevig}}]{moudgalya2020eta}%
  \BibitemOpen
  \bibfield  {author} {\bibinfo {author} {\bibfnamefont {S.}~\bibnamefont
  {Moudgalya}}, \bibinfo {author} {\bibfnamefont {N.}~\bibnamefont
  {Regnault}},\ and\ \bibinfo {author} {\bibfnamefont {B.~A.}\ \bibnamefont
  {Bernevig}},\ }\href@noop {} {\bibfield  {journal} {\bibinfo  {journal}
  {Physical Review B}\ }\textbf {\bibinfo {volume} {102}},\ \bibinfo {pages}
  {085140} (\bibinfo {year} {2020})}\BibitemShut {NoStop}%
\bibitem [{\citenamefont {Pakrouski}\ \emph {et~al.}(2020)\citenamefont
  {Pakrouski}, \citenamefont {Pallegar}, \citenamefont {Popov},\ and\
  \citenamefont {Klebanov}}]{pakrouski2020many}%
  \BibitemOpen
  \bibfield  {author} {\bibinfo {author} {\bibfnamefont {K.}~\bibnamefont
  {Pakrouski}}, \bibinfo {author} {\bibfnamefont {P.~N.}\ \bibnamefont
  {Pallegar}}, \bibinfo {author} {\bibfnamefont {F.~K.}\ \bibnamefont
  {Popov}},\ and\ \bibinfo {author} {\bibfnamefont {I.~R.}\ \bibnamefont
  {Klebanov}},\ }\href@noop {} {\bibfield  {journal} {\bibinfo  {journal}
  {Physical review letters}\ }\textbf {\bibinfo {volume} {125}},\ \bibinfo
  {pages} {230602} (\bibinfo {year} {2020})}\BibitemShut {NoStop}%
\bibitem [{\citenamefont {Ren}\ \emph {et~al.}(2021)\citenamefont {Ren},
  \citenamefont {Liang},\ and\ \citenamefont {Fang}}]{ren2021quasisymmetry}%
  \BibitemOpen
  \bibfield  {author} {\bibinfo {author} {\bibfnamefont {J.}~\bibnamefont
  {Ren}}, \bibinfo {author} {\bibfnamefont {C.}~\bibnamefont {Liang}},\ and\
  \bibinfo {author} {\bibfnamefont {C.}~\bibnamefont {Fang}},\ }\href@noop {}
  {\bibfield  {journal} {\bibinfo  {journal} {Physical Review Letters}\
  }\textbf {\bibinfo {volume} {126}},\ \bibinfo {pages} {120604} (\bibinfo
  {year} {2021})}\BibitemShut {NoStop}%
\bibitem [{\citenamefont {O'Dea}\ \emph {et~al.}(2020)\citenamefont {O'Dea},
  \citenamefont {Burnell}, \citenamefont {Chandran},\ and\ \citenamefont
  {Khemani}}]{o2020tunnels}%
  \BibitemOpen
  \bibfield  {author} {\bibinfo {author} {\bibfnamefont {N.}~\bibnamefont
  {O'Dea}}, \bibinfo {author} {\bibfnamefont {F.}~\bibnamefont {Burnell}},
  \bibinfo {author} {\bibfnamefont {A.}~\bibnamefont {Chandran}},\ and\
  \bibinfo {author} {\bibfnamefont {V.}~\bibnamefont {Khemani}},\ }\href@noop
  {} {\bibfield  {journal} {\bibinfo  {journal} {Physical Review Research}\
  }\textbf {\bibinfo {volume} {2}},\ \bibinfo {pages} {043305} (\bibinfo {year}
  {2020})}\BibitemShut {NoStop}%
\bibitem [{\citenamefont {Yang}(1989)}]{yang1989eta}%
  \BibitemOpen
  \bibfield  {author} {\bibinfo {author} {\bibfnamefont {C.~N.}\ \bibnamefont
  {Yang}},\ }\href@noop {} {\bibfield  {journal} {\bibinfo  {journal} {Physical
  review letters}\ }\textbf {\bibinfo {volume} {63}},\ \bibinfo {pages} {2144}
  (\bibinfo {year} {1989})}\BibitemShut {NoStop}%
\bibitem [{\citenamefont {Yang}\ and\ \citenamefont
  {Zhang}(1990)}]{yang1990so}%
  \BibitemOpen
  \bibfield  {author} {\bibinfo {author} {\bibfnamefont {C.~N.}\ \bibnamefont
  {Yang}}\ and\ \bibinfo {author} {\bibfnamefont {S.}~\bibnamefont {Zhang}},\
  }\href@noop {} {\bibfield  {journal} {\bibinfo  {journal} {Modern Physics
  Letters B}\ }\textbf {\bibinfo {volume} {4}},\ \bibinfo {pages} {759}
  (\bibinfo {year} {1990})}\BibitemShut {NoStop}%
\bibitem [{\citenamefont {Vafek}\ \emph {et~al.}(2017)\citenamefont {Vafek},
  \citenamefont {Regnault},\ and\ \citenamefont
  {Bernevig}}]{vafek2017entanglement}%
  \BibitemOpen
  \bibfield  {author} {\bibinfo {author} {\bibfnamefont {O.}~\bibnamefont
  {Vafek}}, \bibinfo {author} {\bibfnamefont {N.}~\bibnamefont {Regnault}},\
  and\ \bibinfo {author} {\bibfnamefont {B.~A.}\ \bibnamefont {Bernevig}},\
  }\href@noop {} {\bibfield  {journal} {\bibinfo  {journal} {SciPost Physics}\
  }\textbf {\bibinfo {volume} {3}},\ \bibinfo {pages} {043} (\bibinfo {year}
  {2017})}\BibitemShut {NoStop}%
\bibitem [{\citenamefont {Mark}\ and\ \citenamefont
  {Motrunich}(2020)}]{Mark20}%
  \BibitemOpen
  \bibfield  {author} {\bibinfo {author} {\bibfnamefont {D.~K.}\ \bibnamefont
  {Mark}}\ and\ \bibinfo {author} {\bibfnamefont {O.~I.}\ \bibnamefont
  {Motrunich}},\ }\href {https://doi.org/10.1103/PhysRevB.102.075132}
  {\bibfield  {journal} {\bibinfo  {journal} {Phys. Rev. B}\ }\textbf {\bibinfo
  {volume} {102}},\ \bibinfo {pages} {075132} (\bibinfo {year}
  {2020})}\BibitemShut {NoStop}%
\bibitem [{\citenamefont {Pakrouski}\ \emph {et~al.}(2021)\citenamefont
  {Pakrouski}, \citenamefont {Pallegar}, \citenamefont {Popov},\ and\
  \citenamefont {Klebanov}}]{Pakrouski21}%
  \BibitemOpen
  \bibfield  {author} {\bibinfo {author} {\bibfnamefont {K.}~\bibnamefont
  {Pakrouski}}, \bibinfo {author} {\bibfnamefont {P.~N.}\ \bibnamefont
  {Pallegar}}, \bibinfo {author} {\bibfnamefont {F.~K.}\ \bibnamefont
  {Popov}},\ and\ \bibinfo {author} {\bibfnamefont {I.~R.}\ \bibnamefont
  {Klebanov}},\ }\href {https://doi.org/10.1103/PhysRevResearch.3.043156}
  {\bibfield  {journal} {\bibinfo  {journal} {Phys. Rev. Research}\ }\textbf
  {\bibinfo {volume} {3}},\ \bibinfo {pages} {043156} (\bibinfo {year}
  {2021})}\BibitemShut {NoStop}%
\bibitem [{\citenamefont {Jepsen}\ \emph {et~al.}(2022)\citenamefont {Jepsen},
  \citenamefont {Lee}, \citenamefont {Lin}, \citenamefont {Dimitrova},
  \citenamefont {Margalit}, \citenamefont {Ho},\ and\ \citenamefont
  {Ketterle}}]{jepsen2022long}%
  \BibitemOpen
  \bibfield  {author} {\bibinfo {author} {\bibfnamefont {P.~N.}\ \bibnamefont
  {Jepsen}}, \bibinfo {author} {\bibfnamefont {Y.~K.~E.}\ \bibnamefont {Lee}},
  \bibinfo {author} {\bibfnamefont {H.}~\bibnamefont {Lin}}, \bibinfo {author}
  {\bibfnamefont {I.}~\bibnamefont {Dimitrova}}, \bibinfo {author}
  {\bibfnamefont {Y.}~\bibnamefont {Margalit}}, \bibinfo {author}
  {\bibfnamefont {W.~W.}\ \bibnamefont {Ho}},\ and\ \bibinfo {author}
  {\bibfnamefont {W.}~\bibnamefont {Ketterle}},\ }\href@noop {} {\bibfield
  {journal} {\bibinfo  {journal} {Nature Physics}\ }\textbf {\bibinfo {volume}
  {18}},\ \bibinfo {pages} {899} (\bibinfo {year} {2022})}\BibitemShut
  {NoStop}%
\bibitem [{\citenamefont {Matsubara}\ and\ \citenamefont
  {Matsuda}(1956)}]{matsubara1956lattice}%
  \BibitemOpen
  \bibfield  {author} {\bibinfo {author} {\bibfnamefont {T.}~\bibnamefont
  {Matsubara}}\ and\ \bibinfo {author} {\bibfnamefont {H.}~\bibnamefont
  {Matsuda}},\ }\href@noop {} {\bibfield  {journal} {\bibinfo  {journal}
  {Progress of Theoretical Physics}\ }\textbf {\bibinfo {volume} {16}},\
  \bibinfo {pages} {569} (\bibinfo {year} {1956})}\BibitemShut {NoStop}%
\bibitem [{\citenamefont {He}\ and\ \citenamefont {Song}(2025)}]{he2025eta}%
  \BibitemOpen
  \bibfield  {author} {\bibinfo {author} {\bibfnamefont {D.}~\bibnamefont
  {He}}\ and\ \bibinfo {author} {\bibfnamefont {Z.}~\bibnamefont {Song}},\
  }\href@noop {} {\bibfield  {journal} {\bibinfo  {journal} {Physical Review
  B}\ }\textbf {\bibinfo {volume} {112}},\ \bibinfo {pages} {075135} (\bibinfo
  {year} {2025})}\BibitemShut {NoStop}%
\bibitem [{\citenamefont {Zhang}\ and\ \citenamefont
  {Song}(2021{\natexlab{a}})}]{zhang2021topologically}%
  \BibitemOpen
  \bibfield  {author} {\bibinfo {author} {\bibfnamefont {K.}~\bibnamefont
  {Zhang}}\ and\ \bibinfo {author} {\bibfnamefont {Z.}~\bibnamefont {Song}},\
  }\href@noop {} {\bibfield  {journal} {\bibinfo  {journal} {Physical Review
  B}\ }\textbf {\bibinfo {volume} {104}},\ \bibinfo {pages} {184515} (\bibinfo
  {year} {2021}{\natexlab{a}})}\BibitemShut {NoStop}%
\bibitem [{\citenamefont {Wang}\ and\ \citenamefont
  {Song}(2024)}]{wang2024flat}%
  \BibitemOpen
  \bibfield  {author} {\bibinfo {author} {\bibfnamefont {R.}~\bibnamefont
  {Wang}}\ and\ \bibinfo {author} {\bibfnamefont {Z.}~\bibnamefont {Song}},\
  }\href@noop {} {\bibfield  {journal} {\bibinfo  {journal} {Chinese Physics
  Letters}\ }\textbf {\bibinfo {volume} {41}},\ \bibinfo {pages} {047101}
  (\bibinfo {year} {2024})}\BibitemShut {NoStop}%
\bibitem [{\citenamefont {Yang}\ and\ \citenamefont
  {Song}(2022)}]{yang2022dynamic}%
  \BibitemOpen
  \bibfield  {author} {\bibinfo {author} {\bibfnamefont {X.}~\bibnamefont
  {Yang}}\ and\ \bibinfo {author} {\bibfnamefont {Z.}~\bibnamefont {Song}},\
  }\href@noop {} {\bibfield  {journal} {\bibinfo  {journal} {Physical Review
  B}\ }\textbf {\bibinfo {volume} {105}},\ \bibinfo {pages} {195132} (\bibinfo
  {year} {2022})}\BibitemShut {NoStop}%
\bibitem [{\citenamefont {Zhang}\ and\ \citenamefont
  {Song}(2021{\natexlab{b}})}]{zhang2021eta}%
  \BibitemOpen
  \bibfield  {author} {\bibinfo {author} {\bibfnamefont {X.}~\bibnamefont
  {Zhang}}\ and\ \bibinfo {author} {\bibfnamefont {Z.}~\bibnamefont {Song}},\
  }\href@noop {} {\bibfield  {journal} {\bibinfo  {journal} {Physical Review
  B}\ }\textbf {\bibinfo {volume} {103}},\ \bibinfo {pages} {235153} (\bibinfo
  {year} {2021}{\natexlab{b}})}\BibitemShut {NoStop}%
\bibitem [{\citenamefont {Zhang}\ and\ \citenamefont
  {Song}(2020)}]{zhang2020dynamical}%
  \BibitemOpen
  \bibfield  {author} {\bibinfo {author} {\bibfnamefont {X.}~\bibnamefont
  {Zhang}}\ and\ \bibinfo {author} {\bibfnamefont {Z.}~\bibnamefont {Song}},\
  }\href@noop {} {\bibfield  {journal} {\bibinfo  {journal} {Physical Review
  B}\ }\textbf {\bibinfo {volume} {102}},\ \bibinfo {pages} {174303} (\bibinfo
  {year} {2020})}\BibitemShut {NoStop}%
\bibitem [{\citenamefont {Li}\ \emph {et~al.}(2020)\citenamefont {Li},
  \citenamefont {Golez}, \citenamefont {Werner},\ and\ \citenamefont
  {Eckstein}}]{li2020eta}%
  \BibitemOpen
  \bibfield  {author} {\bibinfo {author} {\bibfnamefont {J.}~\bibnamefont
  {Li}}, \bibinfo {author} {\bibfnamefont {D.}~\bibnamefont {Golez}}, \bibinfo
  {author} {\bibfnamefont {P.}~\bibnamefont {Werner}},\ and\ \bibinfo {author}
  {\bibfnamefont {M.}~\bibnamefont {Eckstein}},\ }\href@noop {} {\bibfield
  {journal} {\bibinfo  {journal} {Physical Review B}\ }\textbf {\bibinfo
  {volume} {102}},\ \bibinfo {pages} {165136} (\bibinfo {year}
  {2020})}\BibitemShut {NoStop}%
\bibitem [{\citenamefont {Kaneko}\ \emph {et~al.}(2020)\citenamefont {Kaneko},
  \citenamefont {Yunoki},\ and\ \citenamefont {Millis}}]{kaneko2020charge}%
  \BibitemOpen
  \bibfield  {author} {\bibinfo {author} {\bibfnamefont {T.}~\bibnamefont
  {Kaneko}}, \bibinfo {author} {\bibfnamefont {S.}~\bibnamefont {Yunoki}},\
  and\ \bibinfo {author} {\bibfnamefont {A.~J.}\ \bibnamefont {Millis}},\
  }\href@noop {} {\bibfield  {journal} {\bibinfo  {journal} {Physical Review
  Research}\ }\textbf {\bibinfo {volume} {2}},\ \bibinfo {pages} {032027}
  (\bibinfo {year} {2020})}\BibitemShut {NoStop}%
\bibitem [{\citenamefont {Kaneko}\ \emph {et~al.}(2019)\citenamefont {Kaneko},
  \citenamefont {Shirakawa}, \citenamefont {Sorella},\ and\ \citenamefont
  {Yunoki}}]{kaneko2019photoinduced}%
  \BibitemOpen
  \bibfield  {author} {\bibinfo {author} {\bibfnamefont {T.}~\bibnamefont
  {Kaneko}}, \bibinfo {author} {\bibfnamefont {T.}~\bibnamefont {Shirakawa}},
  \bibinfo {author} {\bibfnamefont {S.}~\bibnamefont {Sorella}},\ and\ \bibinfo
  {author} {\bibfnamefont {S.}~\bibnamefont {Yunoki}},\ }\href@noop {}
  {\bibfield  {journal} {\bibinfo  {journal} {Physical review letters}\
  }\textbf {\bibinfo {volume} {122}},\ \bibinfo {pages} {077002} (\bibinfo
  {year} {2019})}\BibitemShut {NoStop}%
\bibitem [{\citenamefont {He}(2025)}]{dataset}%
  \BibitemOpen
  \bibfield  {author} {\bibinfo {author} {\bibfnamefont {D.~K.}\ \bibnamefont
  {He}},\ }\bibfield  {journal} {\bibinfo  {journal} {Zenodo}\ }\href
  {https://doi.org/http://doi.org/10.5281/zenodo.17394071}
  {http://doi.org/10.5281/zenodo.17394071} (\bibinfo {year} {2025})\BibitemShut
  {NoStop}%
\end{thebibliography}%

\end{document}